\documentclass[prb,showpacs,unsortedaddress,twocolumn,floatfix]{revtex4}

\usepackage{graphicx}
\usepackage{amsmath,amsfonts,amssymb}

\begin{document}

  \bibliographystyle{prsty}

  \title{Unoccupied states of individual silver clusters and chains on Ag(111)}
  \author{A.\ Sperl}
  \author{J.\ Kr\"{o}ger}
  \email{kroeger@physik.uni-kiel.de}
  \author{N.\ N\'{e}el}
  \author{H.\ Jensen}
  \author{R.\ Berndt}
  \affiliation{Institut f\"{u}r Experimentelle und Angewandte Physik, Christian-Albrechts-Universit\"{a}t zu Kiel, D-24098 Kiel, Germany}
  \author{A.\ Franke}
  \author{E.\ Pehlke}
  \affiliation{Institut f\"{u}r Theoretische Physik und Astrophysik, Christian-Albrechts-Universit\"{a}t zu Kiel, D-24098 Kiel, Germany}

  \begin{abstract}
    Size-selected silver clusters on Ag(111) were fabricated with the tip of
    a scanning tunneling microscope. Unoccupied electron resonances give rise to
    image contrast and spectral features which shift toward the Fermi level
    with increasing cluster size. Linear assemblies exhibit higher resonance
    energies than equally sized compact assemblies. Density functional theory
    calculations reproduce the observed energies and enable an assignment
    of the resonances to hybridized atomic $5s$ and $5p$ orbitals with
    silver substrate states.
  \end{abstract}


  \maketitle

  \section{Introduction}
  Metal clusters at the nanometer scale which are supported by surfaces or
  thin films are currently of significant interest. Transport properties,
  \cite{nni_02} catalytic efficiency \cite{sab_00} and selectivity, \cite{sab_01}
  as well as magnetic response \cite{jba_05} depend strongly on the
  size of these assemblies. Moreover, understanding the influence of the substrate
  on the electronic structure of clusters \cite{awa_04} is important
  for new cluster-based materials with tailored optical, catalytic, or magnetic
  properties. To this end, clusters may be decoupled from a metal surface
  by introducing an oxide thin film between the substrate and the
  deposited clusters. \cite{rmj_91,uba_92,cbe_98,mba_99,kho_01}

  Confinement of electrons to a small region leads to the formation of a
  discrete spectrum of their eigenstates. In condensed matter physics,
  spectroscopic studies of the discrete spectrum of individual samples has
  been rather demanding owing to difficulties in preparing and addressing
  suitable single particles. A major step forward was achieved with micro-fabrication
  techniques used to engineer semiconductor quantum dots, the levels of which
  could be resolved in a low temperature range by single-electron tunneling
  spectroscopy. Metals, however, owing to Fermi wavelengths which are of the
  order of a few tenths of a nanometer require extremely small particle sizes
  down to atomic-scale dimensions in order to render quantization phenomena
  observable. These sizes can be routinely obtained nowadays.
  \cite{dcr_95,dcr_01} Besides the required small cluster sizes the fabrication
  of assemblies with a narrow size distribution is another challenging experimental
  task. Several approaches to this goal are known, for instance, the manipulation
  of thermodynamic parameters which dictate a particular growth mode.
  \cite{jav_84,mco_89,rku_90,hbr_98} While lithography or etching-based
  fabrication are exceptionally challenging techniques \cite{pmo_01}
  buffer-layer-assisted growth was shown to lead to particularly narrow size
  distributions of deposited clusters. \cite{lhu_98} The "soft-landing" of
  mass-selected clusters requires a complex apparatus \cite{khm_00} and
  faces the possibility of fragmentation, morphological changes and diversity.
  \cite{sme_00}

  Early scanning tunneling microscopy (STM) imaging of gold and silver clusters
  was reported by Abraham {\it et al.}\,\cite{dwa_86} and Sattler.\cite{ksa_91}
  Pioneering scanning tunneling spectroscopy studies of nanometer-sized
  clusters of gold \cite{rmf_89} and iron \cite{pnf_89} on GaAs(110) and of
  size-selected Si$_{10}$ on Au(100) \cite{yku_89} have been published already
  in 1989. Formation of states in the GaAs band gap was observed in the former
  cases while a variety of cluster images were obtained in the latter case
  despite the deposition of size-selected clusters. An important step forward
  was an experiment reported in Ref.\,\onlinecite{hvr_94}. Mass-selected
  clusters of Pt$_n$ and Pd$_n$ with $1\leq n\leq 15$ on Ag(110) were
  investigated using photoelectron spectroscopy. The authors observed
  size-dependent $d$ states at binding energies around $2\,{\rm eV}$ below
  the Fermi energy. The line widths were observed to be size-dependent as
  well. From a comparison with total-energy calculations a chainlike shape
  of the clusters was inferred.
  \begin{figure*}
    \includegraphics[width=175mm,clip=]{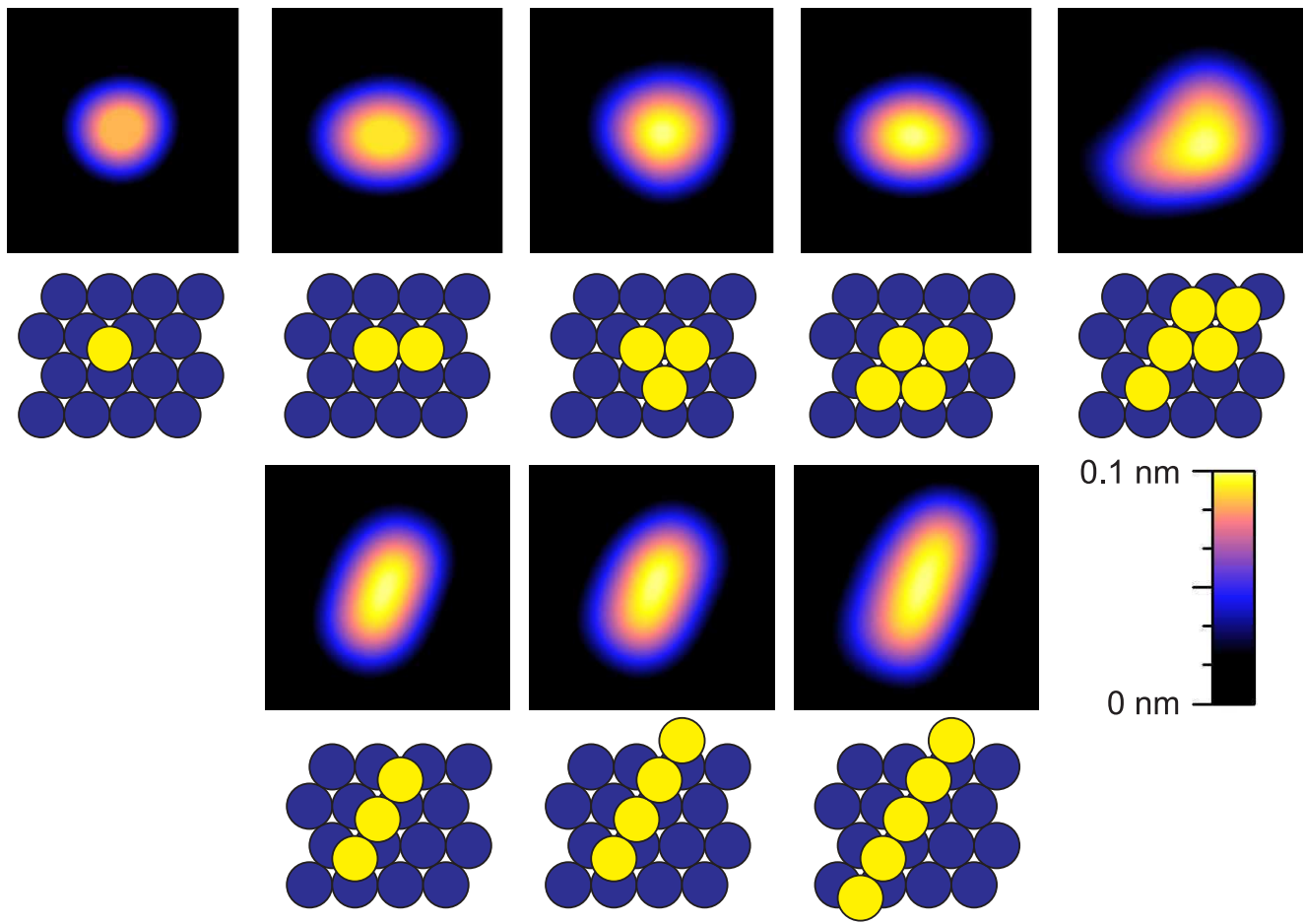}
    \caption[cluster:stm]{(Color online) STM images of silver monomer,
    dimer, trimer, tetramer, and pentamer (from left to right) together with
    sketches of the proposed atomic arrangements (dark and bright circles
    depict substrate atoms and adsorbed atoms, respectively). Sample voltage
    and tunneling current were $V=100\,{\rm mV}$ and $I=0.1\,{\rm nA}$. Image
    sizes: $2.1\,\text{nm}\times 2.1\,\text{nm}$ for Ag$_1$,\ldots,Ag$_5$ and
    $3\,\text{nm}\times 3\,\text{nm}$ for the linear tetramer and pentamer.}
    \label{cluster:stm}
  \end{figure*}

  An ideal experiment enables the control of cluster size and shape
  at the atomic scale. Therefore, atom manipulation with the tip of a
  scanning tunneling microscope was applied in a variety of investigations,
  for instance, gold, manganese, iron, and cobalt dimers on NiAl(100),
  \cite{hjl_04} nickel dimers \cite{vma_02} and chromium trimers
  \cite{tja_01} on Au(111), copper chains on Cu(111), \cite{sfo_04} and
  manganese monomers to tetramers on Ag(111). \cite{jkl_06}
  Quantum confinement of electronic states to chains and islands was also
  revealed for homogeneous metallic systems.
  \cite{sfo_04,jli_98,jla_05,hje_05,scr_05,jkr_05,jkr_07}

  Here we report on studies of electronic properties of silver assemblies
  fabricated by single-atom manipulation on Ag(111) using the tip of a
  low-temperature scanning tunneling microscope. As key results we obtain
  that unoccupied resonances exhibit energies whose actual values
  depend on the size and the shape of the adatom clusters. In particular, linear
  and monatomically wide chains exhibit higher resonance energies
  than their equally sized compact counterparts. Moreover, confinement of the
  unoccupied resonance states to the linear assemblies is found. The energies
  of unoccupied resonances of monomers, dimers, and a long silver
  chain are in agreement with density functional theory calculations.

  \section{Experiment}
  Measurements were performed with a custom-built scanning tunneling microscope
  operated in ultrahigh vacuum at a base pressure of $10^{-9}\,\text{Pa}$ and
  at $7\,\text{K}$. The Ag(111) surface and chemically etched tungsten tips
  were cleaned by argon ion bombardment and annealing. Individual silver atoms
  were deposited onto the sample surface by controlled tip-surface contacts
  as previously described in Ref.\,\onlinecite{lli_05}. Clusters with sizes
  ranging from one to eight atoms were fabricated by tip-induced movements.
  For tunneling resistances of $\approx 10^5\,\Omega$, dragging of
  single silver atoms was feasible. Coalescence of adsorbed atoms (adatoms)
  to dimers up to octamers was accomplished by moving single adatoms close
  enough to the coalescence partner ($\approx$ one nearest-neighbor distance).
  We notice a propensity of silver adatoms to coalesce into compact assemblies
  rather than into linear clusters. For instance, adding an adatom to an already
  existing dimer in most of the cases resulted in a compact trimer rather than
  in a three-adatom chain.
  Silver chains containing more than $100$ atoms were prepared by moving the
  tip toward the surface by $3$ to $5\,\text{nm}$. Various surface dislocations
  were observed to result from this procedure. In particular, extraordinarily
  long and monatomically wide chains were found several hundreds of nanometers
  apart from the indentation area. Spectra of the differential
  conductance ($\text{d}I/\text{d}V$) were acquired by superimposing a
  sinusoidal voltage signal (root-mean-square amplitude $1\,\text{mV}$, frequency
  $10\,\text{kHz}$) onto the tunneling voltage and by measuring the
  current response with a lock-in amplifier. Prior to and in between
  spectroscopy of the clusters the tip status was monitored by giving a sharp onset
  of the Ag(111) surface state band edge in spectra of $\text{d}I/\text{d}V$.
  To obtain sharp onsets of the $\text{d}I/\text{d}V$ signal for the surface
  state and to image single adatoms with nearly circular circumference the tip
  was controllably indented into the substrate. Due to this {\it in vacuo}
  treatment of the tip we expect the tip apex to be covered with substrate
  material. All STM images were acquired in the constant current
  mode with the voltage applied to the sample.

  \section{Theory}
  The total energy of the electronic groundstate and the Kohn-Sham eigenenergies
  have been calculated for the silver monomer and dimer configurations on Ag(111)
  using the Vienna {\it ab initio} simulation package (VASP). \cite{VASP1_93,VASP2_96,VASP3_96}
  Moreover, the Ag chain on Ag(111) has been calculated using the total energy
  package FHI96MD. \cite{fhi_97} Both program packages are based on density
  functional theory with the generalized gradient approximation (GGA)
  (monomer, dimer: PW91 \cite{PW_91}; chain: PBE \cite{PBE_96}) applied to
  the exchange correlation functional. These GGAs are expected to yield
  comparable results. For the monomer and dimer configuration the
  electron-ion interaction is treated within the framework of Bloechl's projector
  augmented wave method (PAW). \cite{PAW_94,PAW2_99} For the calculation of
  the Ag chain a Troullier-Martins pseudopotential has been generated with
  the FHI98PP \cite{PP_98} program. The monomer and dimer configurations have
  been modeled in a slab geometry comprising 14 layers of silver and a
  $(4\times 4)$ or $(5\times 4)$ surface unit cell, respectively. For the
  chain configuration the slab geometry consisted of 14 silver layers and a
  $(9\times 1)$ surface unit cell. Perpendicular to the surface the periodically
  repeated silver slabs are separated by a vacuum region of approximately
  $1.7\,\text{nm}$, which has been the subject to convergence tests and proved
  to be sufficient. In all calculations presented here symmetric slabs were
  chosen such that the adsorption geometry is the same on both sides of the
  slab. The Kohn-Sham wave functions are expanded in a plane wave basis set,
  with a cutoff energy of $250\,\text{eV}$ being sufficient in case of the
  PAW potential. A larger cutoff energy of $544\,\text{eV}$ had to be used
  for the normconserving Troullier-Martins pseudopotential. The integrals over
  the Brillouin zone are approximated by sums over special $k$ points \cite{MP_76}
  using meshes consisting of 16, 9 and 6 $k$ points in the complete first
  Brillouin zone for the monomer, dimer and chain, respectively. The local
  density of states (LDOS) has been calculated using the latter $k$ point
  meshes in case of the monomer and dimer. Additionally, to accurately
  sample the dispersion of the unoccupied state close to the lower one-dimensional
  band edge, a mesh of 144 special $k$ points \cite{MP_76} in the first
  Brillouin zone has been used for the Ag chain. The Kohn-Sham wave functions
  at these additional $k$ points have been calculated via so-called bandstructure
  runs, which are carried out at a frozen electron density from a previous
  self-consistent relaxation. The densities of states have been convoluted
  with a Lorentzian with a full width at half maximum of $150\,\text{meV}$.
  \begin{table}
    \caption[cluster:size]{Apparent heights and full widths at half maximum
    (FWHM) of silver clusters extracted from cross-sectional profiles of STM
    images. The FWHM refers to the longest lateral dimension of linear
    assemblies.}
    \begin{ruledtabular}
      \begin{tabular}{lcc}
        cluster  & height (nm) & FWHM (nm) \\
        \hline
        Ag$_1$   & $0.06$      & $1.06$    \\
        Ag$_2$   & $0.08$      & $1.28$    \\
        Ag$_3$   & $0.10$      & $1.59$    \\
        Ag$_4$   & $0.10$      & $1.72$    \\
        Ag$_5$   & $0.10$      & $2.05$    \\
        Ag$_6$   & $0.10$      & $2.31$    \\
        Ag$_7$   & $0.10$      & $2.69$    \\
        Ag$_8$   & $0.10$      & $3.03$
      \end{tabular}
    \end{ruledtabular}
    \label{cluster:size}
  \end{table}
  Convergence tests for the silver monomer using an 8 layer slab show that
  upon increasing the cutoff energy to $300\,\text{eV}$ the calculated Kohn-Sham
  eigenenergies change by less than $15\,\text{meV}$. Increasing the number
  of $k$ points for the dimer calculation to 16 leads to a change in the Kohn-Sham
  eigenenergies at $\bar{\Gamma}$ of less than $10\,\text{meV}$ (in a test
  calculation for an 8 layer slab).
  \begin{figure}
    \includegraphics[width=85mm,clip=]{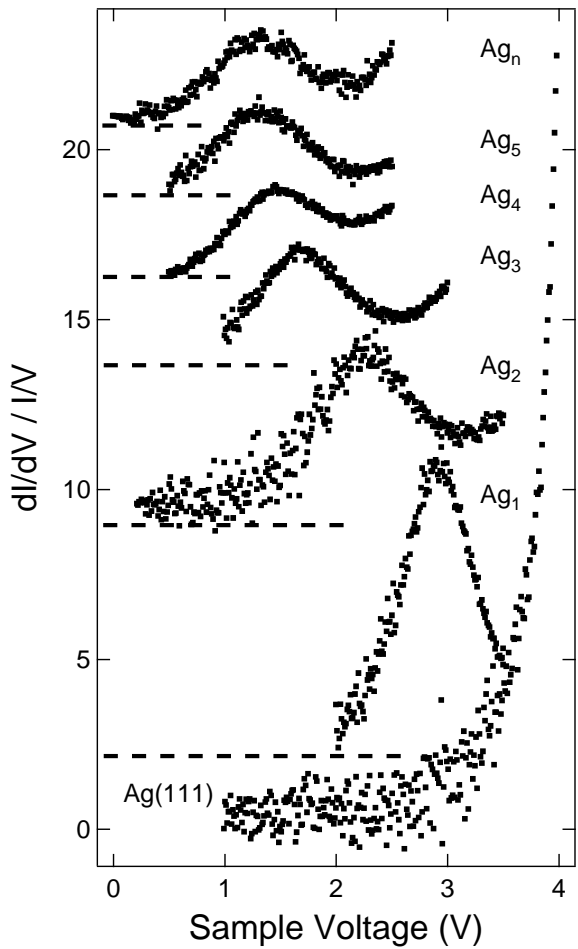}
    \caption[cluster:spec]{Normalized spectra of $\text{d}I/\text{d}V$ acquired
    on clean Ag(111), monomers ($\text{Ag}_1$), dimers ($\text{Ag}_2$), trimers
    ($\text{Ag}_3$), tetramers ($\text{Ag}_4$), and pentamers ($\text{Ag}_5$).
    A compact silver assembly ($\text{Ag}_n$) with probably $n\approx 10$ was
    also analysed. The tunneling gap for the spectra was set at
    $1\,\text{nA}$ and $3.5\,\text{V}$ (Ag$_1$, Ag$_2$), $3.0\,\text{V}$
    (Ag$_3$), $2.5\,\text{V}$ (Ag$_4$, Ag$_5$, Ag$_n$). Spectra of
    $\text{Ag}_n$ $(n\geq 1)$ are vertically offset for clarity. The dashed
    lines indicate the respective zero of the spectra.}
    \label{cluster:spec}
  \end{figure}
  Using the normconserving pseudopotential and the PBE-GGA for exchange and
  correlation, the equilibrium lattice constant of silver is calculated to
  be $0.419\,\text{nm}$. The result is very similar ($0.417\,\text{nm}$) when
  the PAW pseudopotential is used together with the PW91-GGA for exchange
  and correlation. These values are slightly larger than the experimental
  lattice constant of $0.409\,\text{nm}$, but the slight overestimate is
  consistent with other density functional calculations, {\it e.\,g.}, for
  noble metals using GGA functionals. \cite{FBP_98} The slabs were set up
  using the respective theoretical lattice constants. The silver atoms of the
  outermost three layers on both sides of the slab as well as the adatoms were
  allowed to relax without constraints until the residual forces per atom were
  smaller than $7\times 10^{-4}\,\text{Hartree/Bohr}$. The remaining layers
  of the slab were kept fixed at their ideal bulk positions. For the calculation
  of the monomer, one silver adatom is relaxed above the face-centered cubic
  (fcc) hollow site on both sides of the slab, corresponding to a coverage of
  one adatom per 16 surface atoms. For the dimer calculation two silver adatoms
  are relaxed above adjacent fcc hollow sites on both sides of the slab,
  corresponding to a coverage of one dimer per 20 surface atoms. The chain
  geometry consists of silver atoms adsorbed at next-neighbor fcc hollow
  sites in the direction of the chain.

  \section{Results and discussion}
  \subsection{Compact and linear silver clusters: from monomer to octamer}
  Individual clusters with an exactly known number of atoms were fabricated
  by single atom manipulation. The results are presented in
  Fig.\,\ref{cluster:stm}. Compact as well as linear assemblies were produced
  up to sizes of five and eight, respectively (Fig.\,\ref{cluster:stm} shows
  clusters containing five atoms at maximum). We assume that individual
  silver adatoms occupy the threefold coordinated fcc hollow sites of the
  Ag(111) lattice. Table \ref{cluster:size} compares apparent heights and
  full widths at half maximum (FWHM) of linear assemblies with sizes ranging
  from a monomer to an octamer. Cross-sectional profiles of STM images were
  evaluated to this end. Per additional silver atom the length of the
  chains increases by $0.28\,\text{nm}$ on an average, which is in good
  agreement with the nearest-neighbor distance of Ag(111). Starting from the
  trimer the apparent height is $0.10\,\text{nm}$ for all subsequent silver
  assemblies.

  In Ref.\,\onlinecite{jre_03} copper clusters on Cu(111) were investigated
  at $5\,\text{K}$ and the $\text{Cu}_2$ assembly exhibited a nearly circular
  shape in STM images. This observation was attributed to intracell diffusion,
  {\it i.\,e.}, the dimer diffuses within a cell of adjacent hexagonal
  close-packed and fcc sites centered around an on-top site.
  In our case, however, the silver dimer on Ag(111) exhibits different dimensions
  along a close-packed direction ($\approx 1.28\,\text{nm}$) and perpendicular
  to it ($\approx 1.06\,\text{nm}$). We therefore conclude that
  intracell diffusion of a silver dimer adsorbed on Ag(111) plays a minor
  role.

  Next we focus on unoccupied electronic states of the silver assemblies.
  Figure \ref{cluster:spec} shows a series of normalized $\text{d}I/\text{d}V$
  spectra acquired with the tip positioned above the center of compact clusters.
  The tunneling gap for spectroscopy was stabilized at $1\,\text{nA}$ and
  $3.5\,\text{V}$ for Ag$_1$, Ag$_2$, $3.0\,\text{V}$ for Ag$_3$, and
  $2.5\,\text{V}$ for Ag$_4$, Ag$_5$, Ag$_n$. Due to different tip-cluster
  distances for the various spectra we normalized the
  $\text{d}I/\text{d}V$ data sets by the conductance $I/V$ according to
  Refs.\,\onlinecite{rmf_87,ndl_86,vau_96}.
  The spectrum of clean Ag(111) is featureless up to $\approx 3.5\,\text{eV}$.
  A steady increase at higher sample voltages is attributed to field emission
  resonances. \cite{rsb_85,gbi_85} Thus, Ag(111) is a suitable substrate
  for observing unoccupied electronic states of clusters in the range of $0$
  to $\approx 3.5\,\text{eV}$. The $\text{d}I/\text{d}V$ spectrum of a single
  Ag adatom exhibits a pronounced peak slightly below $3\,\text{eV}$.
  \begin{figure}
    \includegraphics[width=85mm,clip=]{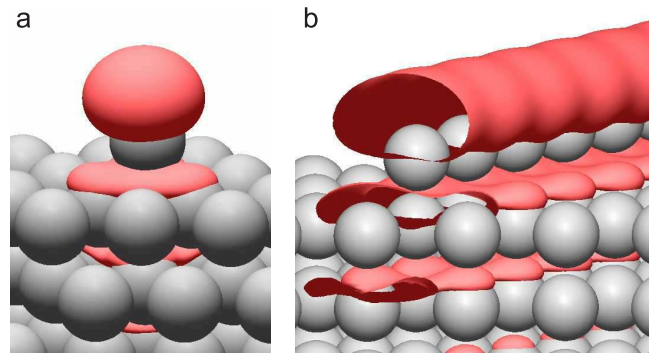}
    \caption[Fig:Wavefunctions]{(Color online) Contour plots of the absolute
    square of the Kohn-Sham wave function at $\bar{\Gamma}$ for the silver
    monomer (a) and the silver chain (b). The corresponding Kohn-Sham
    eigenenergies relative to the Fermi level are $2.66\,\text{eV}$ for the
    Ag monomer and $1.37\,\text{eV}$ for the Ag chain.}
    \label{Fig:Wavefunctions}
  \end{figure}
  By performing spectroscopy in the vicinity of and on the single atom we
  found that the monomer resonance shows a spatial extension comparable to
  the size of the atom in STM images. These results suggest that the silver
  monomer exhibits a quasiatomic resonance. This interpretation is in
  accordance with observations for single Au atoms on NiAl(110) \cite{nni_02}
  and for Pd monomers on $\text{Al}_2\text{O}_3$ layers. \cite{nni03b}
  Thus, the enhanced normalized differential conductance can be attributed to
  resonant tunneling into an empty state of the Ag atom. Indeed, our calculations
  reveal that this state is of {\it sp} character arising from the hybridizaton of
  atomic Ag 5$p_z$ orbitals with 5$s$ admixtures localized at the adsorbate
  and silver substrate states. A typical wave function
  is shown in Fig.\,\ref{Fig:Wavefunctions}a.
  \begin{figure}
    \includegraphics[width=85mm,clip=]{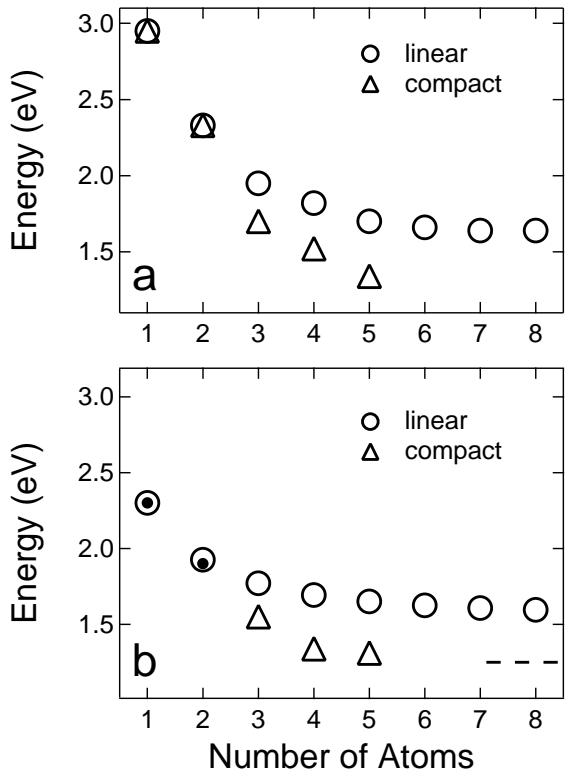}
    \caption[cluster:shift]{Energies of unoccupied resonances as a function
    of cluster size. Resonance energies for compact (triangles) and linear
    (circles) assemblies are presented. (a) Experimental values. Error margins
    for the energies ($\approx\pm 0.05\,\text{eV}$) are the standard deviation
    resulting from a statistical analysis of spectra of a variety of clusters.
    (b) Theoretical values from the tight-binding model described in the text
    (open circles and triangles), compared to {\it ab inito} Kohn-Sham
    eigenenergies (filled circles). The dashed line denotes the lower band
    edge of the dispersion of the chain states as obtained from density
    functional calculations. For each geometry the lowest tight-binding
    eigenenergy is given. The island configurations refer to those displayed
    in Fig.\,\ref{cluster:stm}. The compact islands enfold the trimer, the
    tetramer, and the pentamer.}
    \label{cluster:shift}
  \end{figure}

  Spectra acquired on compact as well as linear clusters containing a higher
  number of atoms likewise exhibit a resonance whose energy shifts to
  lower values with increasing cluster size.
  Figure \ref{cluster:shift}a summarizes the resonance energies for compact
  (triangles) and linear (circles) silver clusters of different sizes. The
  spectra were acquired atop the center of the assemblies.
  The resonance energies for compact clusters are lower than those for their
  equally sized linear counterparts. This observation is in agreement with
  our experience that in the course of atomic manipulation the silver adatoms
  exhibited the propensity to form compact rather than linear clusters. Therefore,
  compact clusters seem to be more stable reflecting the lower energy
  of their resonance. From Fig.\,\ref{cluster:shift}a we further infer that for
  both cluster types the change of the resonance energy becomes less
  pronounced with increasing cluster size. For instance, the energy
  of the compact pentamer resonance is at $\approx 1.5\,\text{eV}$ which
  already comes close to a compact assembly denoted $\text{Ag}_n$ ($n\approx 10$)
  in Fig.\,\ref{cluster:spec}. Lagoute {\it et al.}\,\cite{jla_05} showed for
  Cu adatom islands on Cu(111) an evolution of quasiatomic resonances to the
  two-dimensional Shockley-type surface state. In this study triangular
  Cu adatoms islands containing up to 15 atoms were investigated. In our case,
  an extrapolation of the energy data does not give the binding energy
  of the Ag(111) surface state which is at $\approx -70\,\text{meV}$. We
  propose that this result is due to the shape of our clusters which is
  triangular only for the compact trimer. We will see in the following paragraph
  that a long silver chain exhibits an unoccupied resonance whose energy is
  well above the Fermi level.
  \begin{figure}
    \includegraphics[width=85mm,clip=]{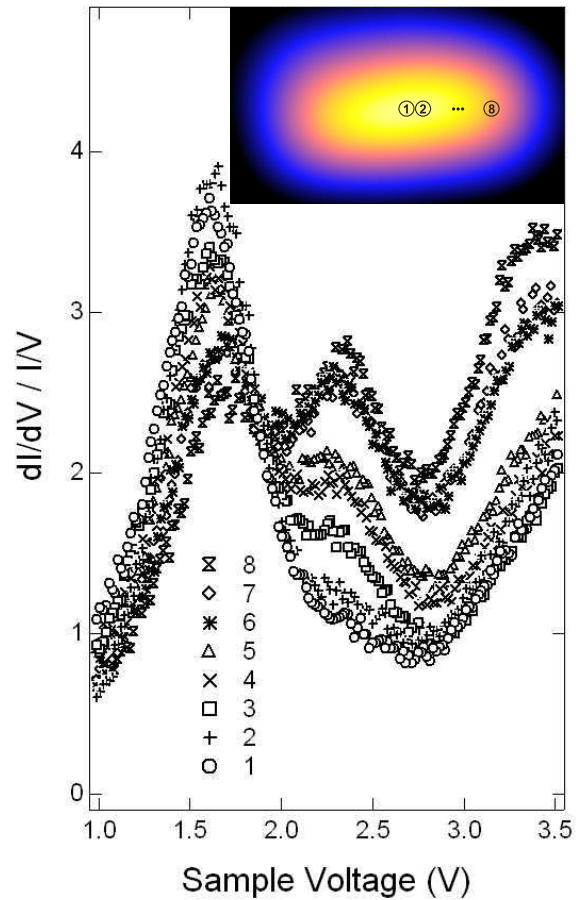}
    \caption[linpent:conf]{(Color online) $\text{d}I/\text{d}V$ spectra acquired
    at indicated (1, 2,\ldots,8) positions of a linear pentamer. Inset: STM
    image of linear pentamer showing the positions at which spectra were
    taken.}
    \label{linpent:conf}
  \end{figure}

  \subsection{Monatomically wide silver chains}
\begin{figure}
    \includegraphics[width=85mm,clip=]{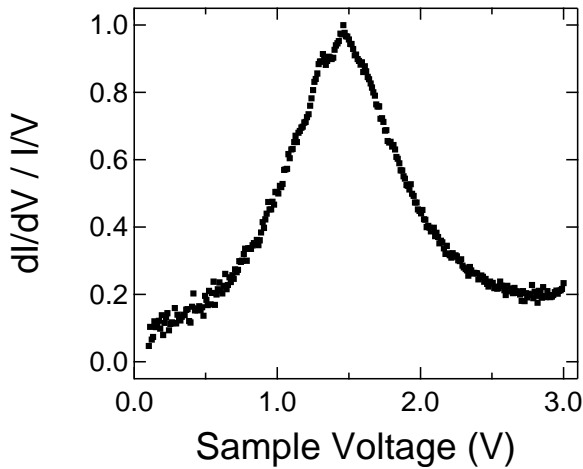}
    \caption[chain:spec]{Normalized spectrum of $\text{d}I/\text{d}V$ at the
    center of a $\approx 45\,\text{nm}$ long and monatomically wide silver
    chain.}
    \label{chain:spec}
  \end{figure}

  One-dimensional metal chains may exhibit interesting properties among
  which the Peierls distortion is probably most famous. \cite{rpp_55,jkr_06}
  This effect describes a modification of the spatial periodicity of the chain
  upon forming an energy gap around the Fermi level. In other words, the
  one-dimensional system gains energy by performing mechanical work for lattice
  deformation and by lowering its electronic energy. For one-dimensional
  silver chains discussed here, however, no changes in the geometric structure
  were observed.
  The adsorption of silver atoms on Ag(111) leads to a hybridization of adatom
  orbitals with substrate electronic states and to resonances whose energies
  are far from the Fermi level (see Figs.\,\ref{cluster:shift} -- \ref{chain:spec}).
  \begin{figure}
    \includegraphics[width=75mm,clip=]{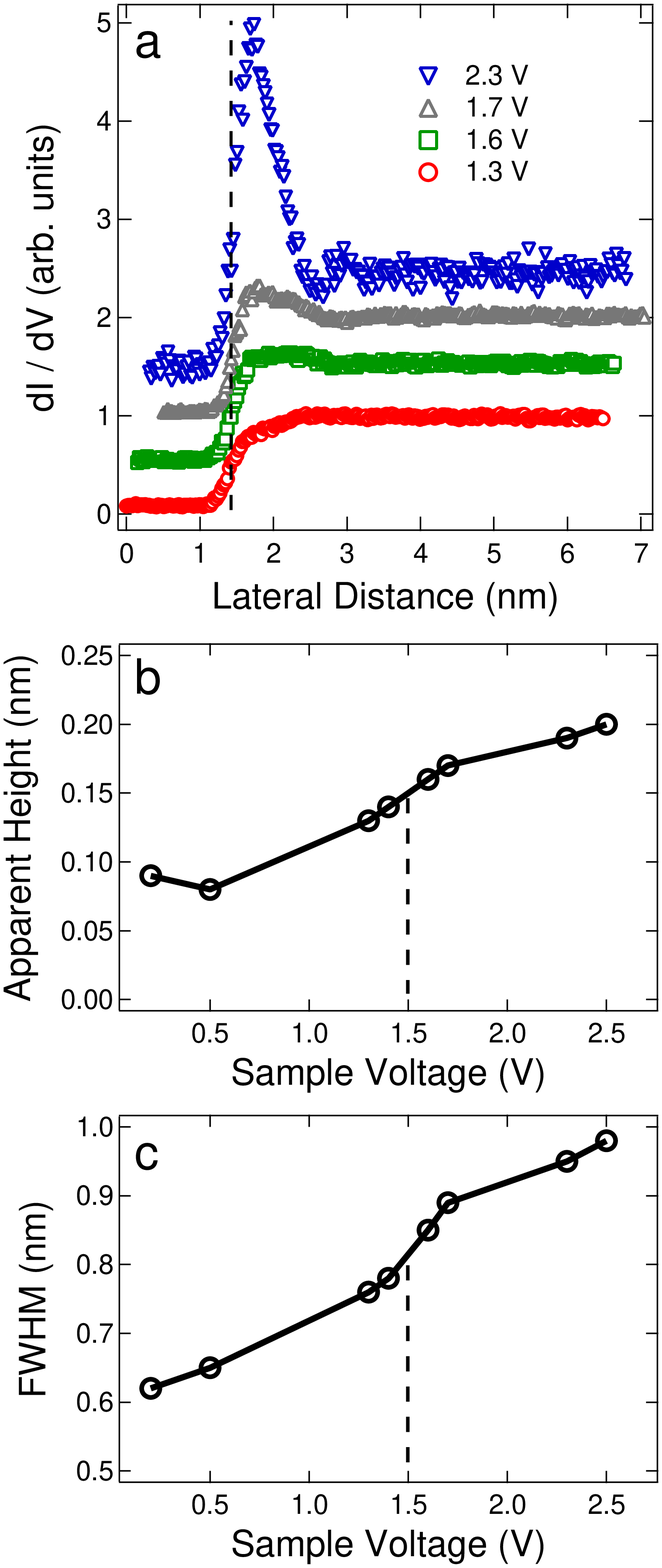}
    \caption[chain:all]{(Color online) (a) Cross-sectional profile in a map
    of $\text{d}I/\text{d}V$ along the middle axis of the chain showing the
    increase of the signal at the end of the chain at a lateral displacement
    of $\approx 1.5\,\text{nm}$. (b) Apparent height of the chain as a
    function of the applied voltage. (c) Like (b) for the full width at half
    maximum of the chain. Dashed lines in (b) and (c) indicate the chain
    resonance energy.}
    \label{chain:all}
  \end{figure}
  Consequently, small deviations in the chain geometric structure are not likely
  to modify the electron occupation of the resonance and therefore are ineffective
  in reducing the total energy of the silver chain. Nevertheless, a different
  combination of substrate and adatom material may lead to resonances close
  to the Fermi level and therefore may favor a Peierls transition.

  Below we focus on electronic properties of monatomically wide silver chains.
  Figure \ref{linpent:conf} shows spatially resolved $\text{d}I/\text{d}V$
  spectra acquired at different sites on a linear pentamer (see inset of
  Fig.\,\ref{linpent:conf}). The $\text{d}I/\text{d}V$ spectrum taken atop
  the center of the assembly exhibits a single peak at $\approx 1.6\,\text{eV}$
  (to be compared also with Fig.\,\ref{cluster:shift}). By moving the
  location of spectroscopy toward the boundary of the linear cluster (spectra
  2,\ldots,8) a gradual trade of spectral weight from the resonance peak at
  $\approx 1.6\,\text{eV}$ to an additional peak at $\approx 2.3\,\text{eV}$
  is observed. Moreover, compared to the resonance energy of
  $\approx 1.6\,\text{eV}$ observed atop the middle of the chain, at its ends
  this energy has shifted up to $\approx 1.7\,\text{eV}$. We attribute the
  presence of an additional peak at the ends of the chain to confinement of
  the resonance to the linear cluster. Similar confinement effects for unoccupied
  resonances were observed for Au chains on NiAl(110) \cite{nni_02} and for
  Cu chains on Cu(111). \cite{sfo_04}

\begin{figure}
    \includegraphics[width=85mm,clip=]{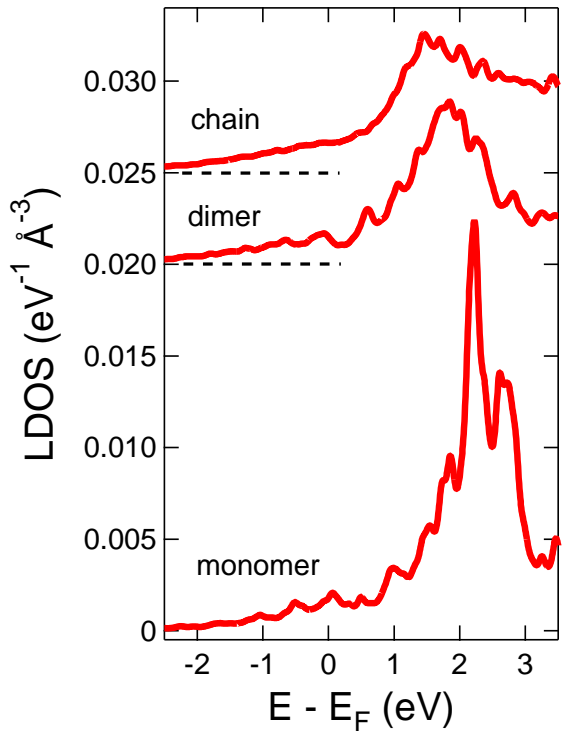}
    \caption[Fig:LDOS]{(Color online) Calculated local density of states
    (LDOS) for (a) a silver monomer, (b) a silver dimer, and (c) an infinitely
    long silver chain on Ag(111). The LDOS has been computed atop a silver
    adatom in case of the monomer and the chain, and atop the center of the
    silver dimer. The peaks are located at $2.3\,\text{eV}$, $1.9\,\text{eV}$,
    and at around $1.5$ -- $1.6\,\text{eV}$, respectively. Calculated LDOS
    for the dimer and the chain were shifted vertically with dashed lines
    indicating the respective zero LDOS.}
    \label{Fig:LDOS}
  \end{figure}

  How does the confinement evolve for chains containing an extremely high number
  of atoms? Starting from the linear octamer it became difficult to resolve
  confinement-related peaks in $\text{d}I/\text{d}V$ spectra, which may be related
  to overlap of neighboring peaks. Nevertheless, confinement was evidenced
  by localization of density of states at the ends of an extremely long chain,
  to be discussed next. The length of the chain is $\approx 45\,\text{nm}$ and
  follows a close-packed direction of the hosting Ag(111) lattice. Consequently,
  the number of silver atoms is approximately $160$. Figure \ref{chain:spec}
  shows a normalized $\text{d}I/\text{d}V$ spectrum of the resonance
  in the middle of the chain. We extract an energy of
  $\approx 1.5\,\text{eV}$ and a FWHM of $\approx 0.6\,\text{eV}$. Our DFT
  calculations reproduce the energy of this chain resonance (see an
  illustration of the wave function in Fig.\,\ref{Fig:Wavefunctions}b and
  compare Figs.\,\ref{chain:spec} and \ref{Fig:LDOS}).

  In addition to the unoccupied resonance, STM images and spatial maps of
  $\text{d}I/\text{d}V$ acquired at different voltages evidence confinement
  of the resonance within the chain. In Fig.\,\ref{chain:all}a spatially
  resolved $\text{d}I/\text{d}V$ data acquired along the long symmetry axis
  of the chain at the indicated voltages are presented. Starting from a voltage
  which corresponds approximately to the resonance energy the
  $\text{d}I/\text{d}V$ signal is piled up at the end of the chain whose
  position is indicated by a dashed line in Fig.\,\ref{chain:all}a).
  In Figs.\,\ref{chain:all}b and \ref{chain:all}c we show the evolution of
  the apparent height and the width (FWHM) of the chain as a function of the
  applied voltage. Both properties exhibit an increase with increasing
  voltage. At sample voltages between $\approx 1.5$ and $\approx 1.7\,\text{V}$
  the apparent height and the width increase stronger than for other voltages.
  Since the chain resonance is located at $\approx 1.5\,\text{eV}$ we
  attribute the higher slopes of the curves to the onset of the resonance.
  The concomitant enhanced density of states may be responsible for the
  observed stronger increase of the apparent height and the FWHM close to the
  resonance energy.

  \subsection{Theoretical results}
  To compare experimental resonance energies with calculated results,
  we evaluated the LDOS at a position of $\approx 0.25\,\text{nm}$ atop the
  adsorbed Ag atom for the monomer and the chain. For the dimer configuration
  the LDOS was computed at approximately the same height atop the center of
  the dimer. The results are presented in Fig.\,\ref{Fig:LDOS}. A similar LDOS
  for Cu chains on Cu(111)  has been calculated by Stepanyuk {\it et al.}
  \cite{VSS05}

  In case of the monomer, a resonance predominantly derived from Ag $sp_z$
  orbitals occurs at $\approx 2.3\,\text{eV}$ above the Fermi energy (see
  Fig.\,\ref{Fig:Wavefunctions}a). In case of the silver dimer, with Ag atoms
  occupying neighboring fcc sites, this resonance splits into a $p_z$ bonding
  resonance at $\approx 1.9\,\text{eV}$ and a $p_z$ antibonding resonance.
  \cite{asp_07} The LDOS of the Ag chain is characterized by a one-dimensional
  band formed from an uncoccupied $p_z$-like resonance as shown in
  Fig.\,\ref{Fig:Wavefunctions}b. The lower band edge of this one-dimensional
  band is located at $\approx 1.3\,\text{eV}$ above the Fermi energy $E_{\text{F}}$
  (as derived from the electronic eigenenergies at $\bar{\Gamma}$). A peak
  arises at around \mbox{1.5 -- 1.6\,\text{eV}}. No upper band edge of the
  one-dimensional band was observed for energies below the work function of
  silver. Table \ref{comp} summarizes experimental and calculated resonance
  energies. Owing to the agreement between experiment and theory we interpret
  peaks in the $\text{d}I/\text{d}V$ spectra as the signature of $sp_z$- or
  $p_z$-like Ag adsorbate resonances and their electronic interaction with
  silver substrate states.
  \begin{table}
    \caption[comp]{Comparison of experimental (exp) and calculated (cal)
    energies ($E$) of unoccupied resonances.}
    \begin{ruledtabular}
      \begin{tabular}{lcc}
        cluster  & $E^{\text{exp}}$ (eV) & $E^{\text{cal}}$ (eV) \\
        \hline
        monomer  & $2.9$                 & $2.3$                 \\
        dimer    & $2.3$                 & $1.9$                 \\
        chain    & $1.5$                 & $1.5$ -- $1.6$
      \end{tabular}
    \end{ruledtabular}
    \label{comp}
  \end{table}

  We did not perform {\it ab initio} calculations for silver clusters
  of sizes larger than two atoms due to the large computational costs arising
  from the increasing size of the surface unit cell. However, we provide
  estimates for the electronic eigenenergies of larger clusters by means of
  a simple tight-binding model. The purpose of this estimate is to explain
  the energy shifts observed by tunneling spectroscopy semi-quantitatively.
  In our tight-binding approach the substrate is not considered explicitly,
  {\it i.\,e.}, the islands are represented by free-standing two-dimensional
  clusters. We include one Ag $sp_z$ orbital per atom. There are only two
  free tight-binding parameters: the orbital energy $\varepsilon_0$ and the
  next-neighbor transfer matrix-element $t$, which accounts for the direct
  interaction between nearest neighbor Ag atoms and, implicitly, part of the
  interaction via the Ag substrate. All further interactions with respect
  to more distant atoms are neglected, as is the variation of the crystal-field
  energy shift of the orbital energy for different geometrical environments.
  As usual, orbital overlaps are not accounted for explicitly.

  The tight-binding parameters are consistently derived from our DFT results,
  {\it i.\,e.}, the resonance energy of the monomer $\varepsilon_0 = 2.3\,\text{eV}$
  (experimental value $2.9\,\text{eV}$) and the binding energy of the Ag(111)
  surface state at $\bar{\Gamma}$ $\varepsilon_0 + 6t = +0.05\,\text{eV}$
  (experimental value $-0.07\,\text{eV}$)\cite{LSB:97} are reproduced by the
  tight-binding model. The quality of the tight-binding results can be estimated
  from comparison with the DFT Kohn-Sham eigenenergies of the dimer and the chain
  shown in Fig.\,\ref{cluster:shift}b.
  The lowest energy eigenvalue is given for each configuration. For further
  evaluation of the quality of the tight-binding model we notice that for the
  effective mass $m^*$ of the surface state we obtain $0.8\,\text{m}_{\text{e}}$
  ($\text{m}_{\text{e}}$ is the free electron mass) from tight-binding
  calculations to be compared with a DFT value of $0.39\,\text{m}_{\text{e}}$
  and an experimental value of $(0.42 \pm 0.02)\,\text{m}_{\text{e}}$.
  \cite{LSB:97} The effective mass of the $sp_z$ resonance at the one-dimensional
  Ag chain is $1.2\,\text{m}_{\text{e}}$ in our tight-binding approach
  to be compared with a value of about $0.6\,\text{m}_{\text{e}}$ derived
  from the dispersion of the Kohn-Sham eigenenergies close to $\bar{\Gamma}$.
  Most probably, the overestimate of the effective mass by a factor of two
  in both cases may be partially due to the fact that no parameter describing
  the crystal-field energy shift is included in the tight-binding Hamilton
  operator giving rise to an inaccurate value of the transfer parameter.

  Nevertheless, the simple tight-binding approach provides all qualitative
  trends for the cluster eigenenergies (see Fig.\,\ref{cluster:shift}b).
  Compact clusters have lower eigenenergies than equally sized linear
  assemblies and the trimer exhibits a lowest electronic eigenenergy which
  is close to the lower band-edge of the infinite chain.

  \section{Summary}
  Size-selected silver clusters were fabricated by tip-assisted single-atom
  manipulation on Ag(111). Unoccupied electronic resonances exhibit energies
  which are characteristic of size and shape of the silver assemblies.
  In particular, the resonances of linear clusters have higher energies than
  the resonances of equally sized compact clusters. For both types of clusters
  the resonance energy shifts toward the Fermi energy with increasing cluster
  size. These observations are qualitatively in agreement with a tight-binding
  model of the clusters. Calculations based on density functional theory
  model the energies of monomers, dimers, and monatomically wide infinitely
  long chains. The resonances are of $sp$ character and arise from Ag $5p_z$
  orbitals (with $5s$ admixtures) which are localized at the adsorbate atom
  and hybridize with silver substrate states.

  Funding of this work by the Deutsche Forschungsgemeinschaft through grant
  number SPP 1153 is gratefully acknowledged.

\end{document}